\documentclass{llncs}

\usepackage[english]{babel}   
\usepackage{hyperref}               
\usepackage{graphicx} 
\usepackage{pgfplots}
\usepackage{pgfplotstable}
\usepackage{multicol}
\usepackage{float}
\usepackage{color}
\usepackage{listings}
\usepackage[style=authoryear,backend=biber]{biblatex}
\bibliography{tipiru}   

\begin{document}

\title{Neural Network-based exploration of construct validity for Russian version of the 10-item Big Five Inventory}
\author{Anastasia Sergeeva\textsuperscript{1} \and Bogdan Kirillov \textsuperscript{2} \and Alyona Dzhumagulova\textsuperscript{1}}
\institute{ITMO University \and Skolkovo Institute of Science and Technology}
\maketitle

\begin{abstract}

This study aims to present a new method of exploring construct validity of questionnaires based on neural network. Using this test we further explore convergent validity for Russian adaptation of TIPI (Ten-Item Personality Inventory by Gosling, Rentfrow, and Swann). Due to small number of questions TIPI-RU can be used as an express-method for surveying large number of people, especially in the Internet-studies. It can be also used with other translations of the same questionnaire in the intercultural studies. The neural network test for construct validity can be used as more convenient substitute for path model.

\end{abstract}

\section{Introduction}\label{sec:Introduction}

\paragraph{}
Questionnaires are a viable tool in modern psychological research but to be useful they have to be validated in a number of ways. Usual questionnaire validation pipeline involves a test for internal consistency, construct validity and reliability (\cite{furr2017psychometrics}). Our research is centered around second issue, construct validity. Construct validity shows how well the questionnaire can generalize, to what degree the measures that the questionnaire provides can be applied to estimate the characteristics of psychological model.
\paragraph{}
Construct validity is categorized into convergent and discriminant validity (\cite{campbell1959convergent}):
	\begin{enumerate}
		\item Convergent validity - "Are two theoretically related ways of estimating the same characteristic really related?";
		\item Discriminant validity - "Are two theoretically unrelated ways of estimating the same characteristic really unrelated?"; 
	\end{enumerate}
Both types of validity are important to investigate the construct validity of a questionnaire. Convergent and discriminant validation ensure that the questionnaire does precisely what it is made to do - provides a way to test a psychological model experimentally.
\paragraph{}
In this study we provide a novel approach to construct validity evaluation. It is based on usage of neural network to predict characteristics of well-established questionnaire from items of the questionnaire under investigation. Using direct prediction we can evaluate convergent validity and we can evaluate discriminant validity using interpretation of trained weights.
\paragraph{}
The Five Factor Model of personality traits (the Big Five, also known as the OCEAN or CANOE model) is currently among the most used personality traits models. Questionnaires based on its factors (Neuroticism, Extraversion, Openness to experience, Agreeableness and Conscientiousness) are widely used in scientific and industrial applications that require personality diagnosis. The major drawback of existing questionnaires is their size - it ranges from 44 items (BFI (\cite{Joh99})) to 240 items (\cite{CM92}). Size makes research more difficult especially in Internet-based cases where lack of outside control (usually provided by researchers in offline cases) and participant's preference to skip tedious tasks condone random response or quitting. 

\paragraph{}
A number of studies (\cite{Cre12}, \cite{Gun15}) suggest 10-item personality questionnaires as brief diagnostic tools for they have satisfying psychometric performance. In this study we choose to use a Russian adaptation of TIPI (\cite{Gos03}) questionnaire due to the TIPI's cross-cultural generalizability (which is shown by a set of TIPI international adaptations (\cite{Her},  \cite{Muc07}, \cite{Hof08}, \cite{Osh12}, \cite{Rom12}, \cite{Ren13}, \cite{Lag14}, \cite{Chi15} etc).

\paragraph{}
Currently there are two competing Russian adaptations for TIPI questionnaire, KOBT (\cite{KT16}) and TIPI-RU (\cite{sergeeva2016translation}). Their performance in convergent validity are close (TIPI-RU performs slightly better in Extraversion, Agreeableness and Emotional stability, KOBT - in Openness and Conscientiousness). For the current study TIPI-RU is considered a better alternative. TIPI-RU data are freely and openly available at github (\cite{TIPIDATA}) so the data can be used as an addition to our own sample. We use TIPI-RU as an example for application of our novel convergent validity evaluation method.

\section{Materials and methods}\label{sec:MatMeth}

\subsection{TIPI-RU questionnaire}
\paragraph{}
The TIPI-RU is translated and validated version of TIPI questionnaire (\cite{Gos03}). The translation can be found in appendix 1 of (\cite{sergeeva2016translation}). Questionnaire consists of 10 questions (below denoted as $TIPI_n$ where $n$ ranges from 1 to 10). The actual big five characteristics are computed from answers according to the following formulae:
\begin{equation}
	\label{tipi-e}
	E = 0.5 (TIPI_1 + reverse(TIPI_6) 
\end{equation}
\begin{equation}
	A = 0.5 (TIPI_7 + reverse(TIPI_2) 
\end{equation}
\begin{equation}
	C = 0.5 (TIPI_3 + reverse(TIPI_8) 
\end{equation}
\begin{equation}
	ES = 0.5 (TIPI_9 + reverse(TIPI_4) 
\end{equation}
\begin{equation}
	\label{tipi-o}
	O = 0.5 (TIPI_5 + reverse(TIPI_{10}) 
\end{equation}
where $reverse$ means taking the opposite value on Likert scale (7 becomes 1, 6 becomes 2 etc.) and big five characteristics are denoted (here and in all plots) by first letters of their corresponding names (Extraversion, Agreeableness, Consciousness, Emotional stability and Openness).

\subsection{Dataset}
We use the extended dataset of 457 observations that include the one composed by Sergeeva, Kirillov and Dzhumagulova (218 observations). 218 old observations are freely available at (\cite{TIPIDATA}) and we have collected other 239 by surveying Russian students who did not participate in the research of Sergeeva's group.
\begin{figure}[!h]
	\centering
	\includegraphics[width=0.7\textwidth]{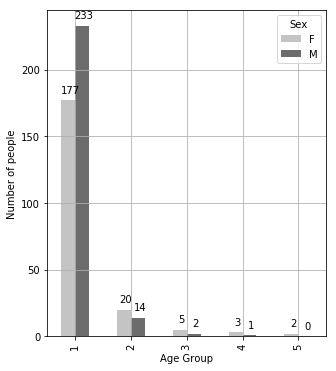}
	\caption{Gender-Age distribution}
	\label{fig:fig0}
\end{figure}
The age groups on the figure \ref{fig:fig0} are: 1 (10 - 19 years), 2 (20 - 29 years), 3 (30 - 39 years), 4 (40 - 49 years) and 5 (50 - 59 years).

\subsection{Convergent validity test via neural network}
\paragraph{}
To check convergent validity of TIPI-RU via neural network we use the following scheme:
\begin{enumerate}
	\item We use 5PFQ as template and we assume that its characteristics (extraversion, emotional stability etc.) can be measured simpler via TIPI-RU questionnaire;
	\item If the latter is correct, we can fit a neural network to predict 5PFQ characteristics from answers to TIPI-RU questions;
	\item For our approach to be successful we have to address the following issues:
	\begin{itemize}
		\item Ensure that the network is learning something, preferrably a certain mapping from TIPI-RU answers to 5PFQ characteristics;
		\item Ensure that even the small network does overfit - it shows that there is a really strong connection between inputs and outputs;
		\item Ensure that the result of trained network is different from results of network trained on random permutations of labels.
	\end{itemize}
	\item Evaluate the model's performance via quality measures.
\end{enumerate}
\paragraph{}
Sergeeva, Kirillov and Dzhumagulova among other methods used path model to confirm convergent validity. The neural network approach is conceptually much simpler: path model is based on fitting five different linear regression models to predict a 5PFQ item given the answers to questions that construct its TIPI-RU counterpart, but with neural network we can predict all five 5PFQ characteristics using all 10 TIPI-RU questions with single model. If the network can do it, then TIPI-RU and 5PFQ converge - it proves the convergent validity of TIPI-RU. The process of TIPI-RU computation can be viewed as application of a single hidden layer neural network with the following structure (for simplicity, we assume that Likert reverses of answers to 2,4,6,8,10 questions are already taken before the application of the network):
 
\begin{figure}[!h]
	\centering
	\includegraphics[width=\textwidth]{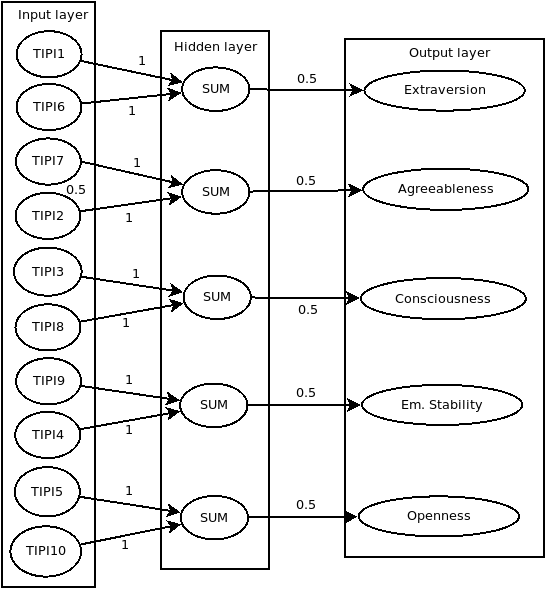}
	\caption{TIPI computation as neural network. Only non-zero weights are shown.}
	\label{fig:fig1}
\end{figure}
\paragraph{}
SUM here denotes output of a summarizing neuron with linear activation function and, for simplicity, this toy network has no bias, so the following is correct: 
\begin{equation}
	SUM = W \dot X
\end{equation}
where $W$ is vector of weights, and $X$ is the vector of inputs to this particular neuron. 
\paragraph{}
Same operation is performed at the output layer neurons. The network is fully connected but all edges that don't add up to TIPI-RU scales are set to zero. They are not shown on figure \ref{fig:fig1}.

\paragraph{}
This particular configuration is very hard to reach by gradient descent. It is not impossible, but very improbable for a network to converge there. But the network can converge to a different weight set that allows for non-zero edges that connect output characteristics and questions that aren't included in them. We don't really need to follow the template set by figure \ref{fig:fig1} and can use an arbitrary neural network. Any network that is reasonably small should be enough. 

\paragraph{}
To ensure that overfitting during the training may be attributed to strong connections between inputs and outputs rather than to model's power we should keep the model as small as possible. There is a trade-off between susceptibility to overfitting and ability to fit anything useful: number of parameters should be large enough to recover a dependency between inputs and outputs but at the same time small enough but it should be still pretty small because the network should find it hard to remember every observation training and validation set. Usually the penalty on model size is considered an auxilliary way of regularization and primarily other ways like L1/L2 regularization are used. For this case we find convenient to use the size as regularizer only.

\paragraph{}
In the current study we use the following configuration of network:
\begin{figure}[H]
	\centering
	\includegraphics[width=0.4\textwidth]{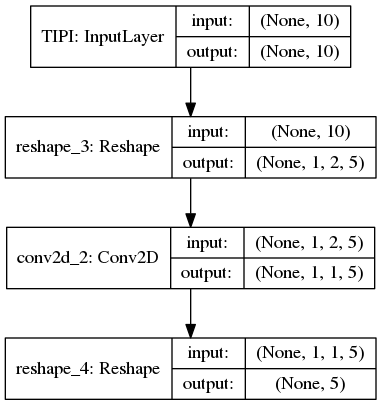}
	\caption{Actual network that learns TIPI-5PFQ connection}
	\label{fig:fig2}
\end{figure}
\paragraph{}
It is a very small convolutional network that consist of one 2D convolutional layer and two reshaping operations. The first reshape is needed to reshape the incoming TIPI items into "pictures" that are 1 in height, 2 in width and have five channels. The second reshape turns (1,1,5) output of convolutional layer into just 5 answers.

\paragraph{}
The network has 50 parameters - exactly two thirds of presumed TIPI network shown at figure \ref{fig:fig1} that has 75. A pretty simple way to see whether the network is learning something correct and it is not by happenstance is to destroy any real structure that is present in data, then try to fit the network from such damaged set and compare the distribution of results with ones of networks trained on unharmed data. To do so, we perform an investigation obeying the following scheme:
\begin{enumerate}
	\item Train 100 networks on the TIPI-RU dataset - it will provide distribution of MSEs on unharmed data;
	\item Shuffle 5PFQ characteristics corresponding to TIPI-RU answers at random, then train a network to predict shuffled 5PFQs from same TIPI-RUs.
	\item Do step 2 one hundred times;
	\item Check whether two samples of error measures come from the same distribution or not by two-sample Kolmogorov-Smirnov test.
\end{enumerate}

\paragraph{}
Difference between two distributions shows that there is a true dependency between answers to TIPI questions and 5PFQ characteristics that we got completely destroyed while shuffling the labels. The network is implemented using Keras (\cite{chollet2015keras}) with Tensorflow (\cite{tensorflow2015-whitepaper}) as backend. All plots are made with Matplotlib (\cite{hunter2007matplotlib}). 

\paragraph{}
We make here a reasonable assumption that the best predictions of 5PFQ that a generalizable (not overfitted) model can make from the TIPI-RU data are actually the TIPI-RU values themselves. So we can find a best MSE possible by computing MSE between 5PFQ characteristics and TIPI characteristics both divided by corresponding maximas. A network that gets below that threshold is overfitting. 

\subsection{Neural network performance measures}
\paragraph{}
This work uses classical loss function for regression: mean squared error (MSE).
\begin{equation}
	MSE(\hat{y}, y) = \frac{1}{n}\sum_{i=0}^n(y-\hat{y})^2
\end{equation}
where $y$ - real value, $\hat{y}$ - value predicted by model.
\paragraph{}
We choose MSE as loss function instead of mean absolute error because it punishes large deviations from the real value more than small ones. But we use MAE as an additional performance measure:
\begin{equation}
	MAE(\hat{y}, y) = \frac{1}{n}\sum_{i=0}^n|y-\hat{y}|
\end{equation}

\subsection{Discriminant validity via interpretation of trained weights}
\paragraph{}
Neural Network Interpretation is a complex task and currently a lot of approaches of solving it exists. For a thorough review one can take a look at (\cite{montavon2017methods}). But in this particular case the interpretation becomes very simple.
\paragraph{}
According to definition of convolution (\cite{goodfellow2016deep}), the output of single convolution is as follows:
\begin{equation}
	Z = \sum_{i=1,j=1}^{n,m} X_{i,j} \times W_{i,j},
\end{equation}
where \(W\) - weights of single convolutional filter of interest, \(X\) - single window from the input image.
\paragraph{}
The resulting output of a convolution filter (without activation function) is constructed by applying this operation to whole input example via a sliding window. The important implication of the convolution is that weights of the convolutional layer after the end of training will mimic structure of its input. A set of CNN interpretation methods is based on this property but for our simple case we need only to visualize the weights for each neuron and it will be enough to observe captured structure. Structure similar to figure \ref{fig:fig1} and equations \ref{tipi-e}-\ref{tipi-o} shows high level of discriminant validity.

\section{Results}\label{sec:Results}
The network reaches minimal MSE possible somewhere around 60-th epoch of training. Minimal MSE possible is equal 0.05. Validation set is randomly chosen 40\% of the whole: 
\begin{figure}[H]
	\centering
	\includegraphics[width=\textwidth]{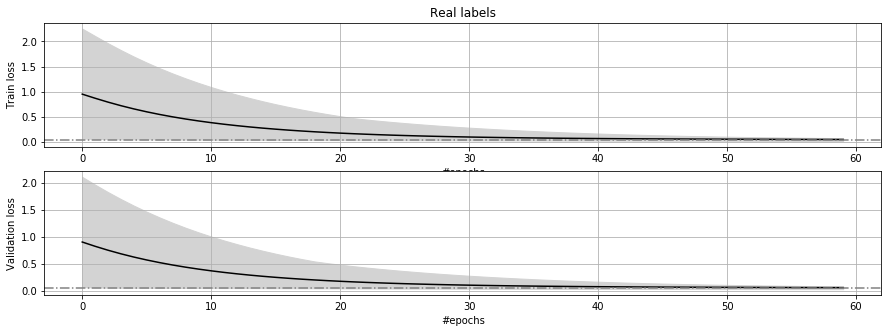}
	\caption{First sixty epochs of training on real labels. Horizontal line denotes minimal MSE possible. Black line is mean of 100 repetitions, grey area is error region.}
	\label{fig:fig3}
\end{figure}

\paragraph{}
After reaching the minimal possible MSE the network overfits and drops MSE to 0. During training the distribution of MSEs and MAEs from models trained on reshuffled labels diverges from one of the models trained on correct labels as shown on figure \ref{fig:fig4}:

\begin{figure}[H]
	\centering
	\includegraphics[width=\textwidth]{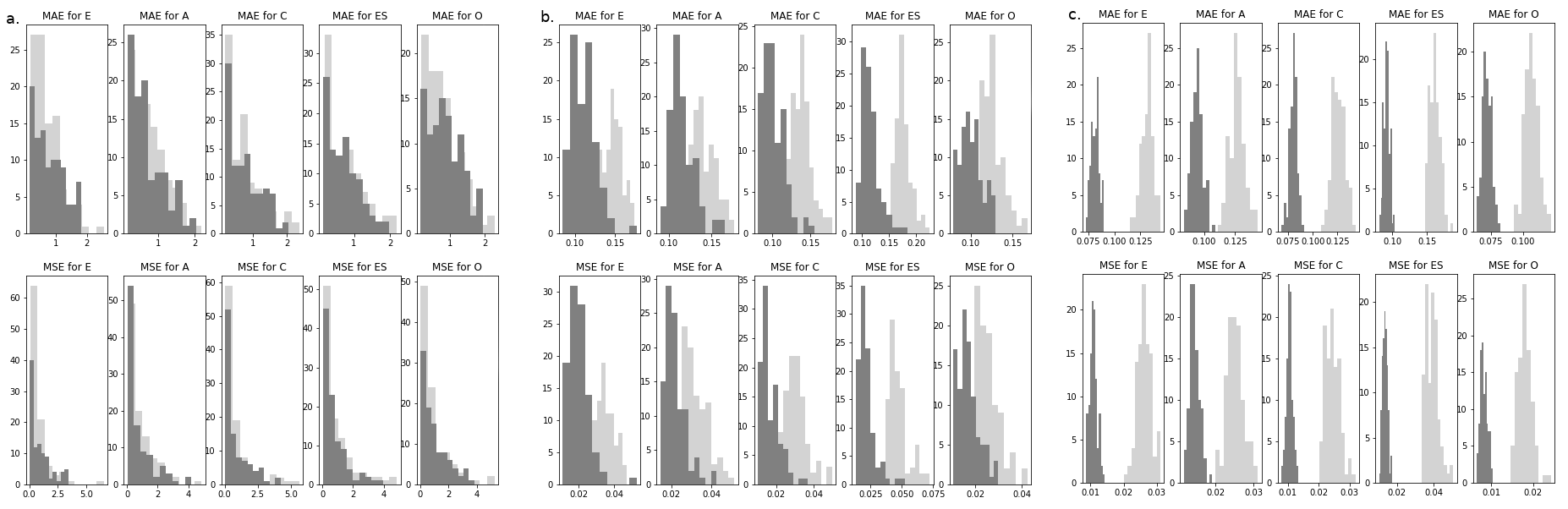}
	\caption{Divergence of correct and permuted label distributions: \textbf{a}. start of training, \textbf{b.} 125th epoch, \textbf{c.} end of training. Dark grey - correct models, light grey - models trained on permuted labels.}
	\label{fig:fig4}
\end{figure}
\paragraph{}
Initially both kinds of models are indistinguishable. But as the training goes, the correct models drive towards zero while permuted ones do not and at the end two distributions are no more the same. The full animation is freely available on YouTube (\cite{Divergence}). It is the visual, qualitative way to check whether there is a strong connection between inputs and outputs. The quantitative one that we use is to apply two-sample Kolmogorov-Smirnov test. For computation we use Scipy KS-test implementation (\cite{jones2014scipy}).

\begin{figure}[H]
	\centering
	\includegraphics[width=\textwidth]{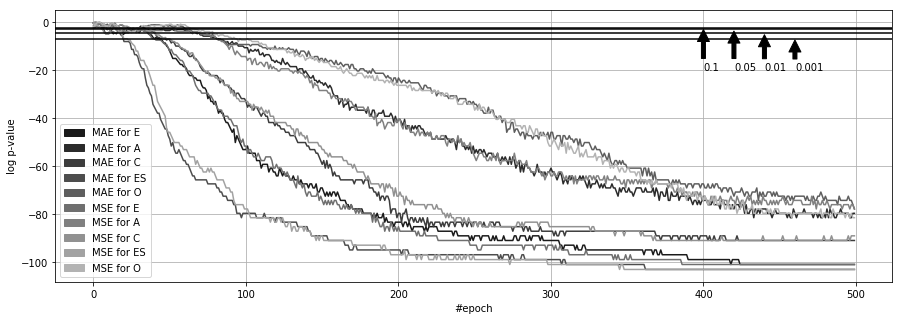}
	\caption{Two-sample KS-test p-values (natural logarithms of them) for MAE and MSE as a function of training time. Black arrows denote different thresholds.}
	\label{fig:fig5}
\end{figure}

\paragraph{}
The question KS-test answers is "Were these two samples drawn from the same distribution?". Null hypothesis is that they are, so for our case p-values should be below the threshold. The differences, as shown on figure \ref{fig:fig5}, grow enough to pass even the most strict threshold quite fast. Weight visualization, as present at figure \ref{fig:fig6}, shows the structure similar to figure \ref{fig:fig1} and equations \ref{tipi-e}-\ref{tipi-o}:

\begin{figure}[H]
	\centering
	\includegraphics[width=0.4\textwidth]{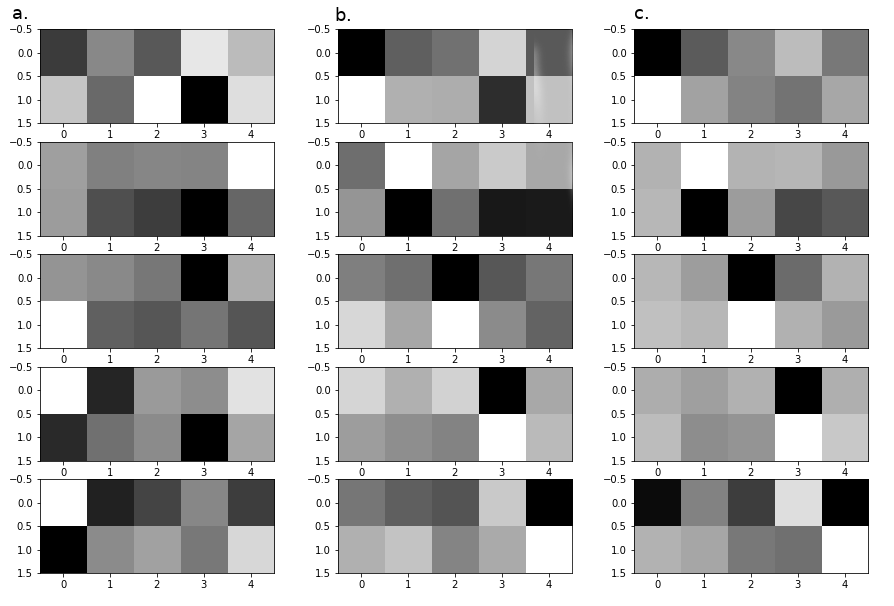}
	\caption{Evolution of weights during training (average over 100 runs): \textbf{a}. start of training, \textbf{b.} 70th epoch, \textbf{c.} end of training. Single image represents a single convolutional neuron. Black - large positive weight, white - large (in absolute value) negative weight, shades of grey - weights close to 0.}
	\label{fig:fig6}
\end{figure}
\paragraph{}
It even captures the sign reversal in Agreableness (second row). Also if one looks closely at the Openness neuron (fifth row), it will be obvious that Openness lacks discriminant validity - there are a lot of other items that are highlighted as strong as the valid column. This inconsistency of Openness was already described by Sergeeva et al. and by other researchers. Neural Network method for validity testing converges with more conventional approaches despite being mostly qualitative.

\section{Conclusion}\label{sec:Conclusion}

\paragraph{}
In this study we have introduced a novel qualitative method for evaluation of construct validity of personality questionnaire using a neural network-based approach. The method is easy to implement with modern Deep Learning frameworks and is more interpretable than traditional methods like path models or correlation matrices since it answers much simpler question: "Is there a learnable connection between inputs (answers to questionnaire) and outputs (characteristics of questionnaire made to measure the same constructs)?". An obvious drawback of our method is that no simple way of comparing its performance to path model exists. 

\paragraph{}
Based on our core findings we can recommend using TIPI-RU as a brief method for measuring personality in non-clinical settings like the internet-assesment of personality measures as it passes the neural network test. Future studies involve using neural network test for evaluation of other questionnaires and exploring its limits of applicability. 

\paragraph{}
All procedures involving human participants  performed in this study went in accordance with the ethical standards of the institutional and/or national research committee and with the 1964 Helsinki declaration and its
later amendments or comparable ethical standards.






\newpage
\printbibliography
\newpage

\end{document}